\documentclass{article}
\usepackage{makeidx}         
\usepackage{graphicx}        
\usepackage{multicol}        
\usepackage[bottom]{footmisc}
\usepackage{bm}      
\renewcommand{\vec}{\mathbf}
\def\neu#1{#1} 

\def\pre{{Phys.~Rev.~E}}        
\def\apj{{Astrophys. Journal}}                 

\title{The effect of subfilter-scale physics on regularization models}
\author{Jonathan Pietarila Graham$^{1}$, Darryl D. Holm$^{2}$, Pablo Mininni$^{3,4}$,\\ and Annick Pouquet$^{3}$}
\date{\today}
\begin{document}
   \maketitle



%
{$^1$Max-Planck-Institut f\"ur Sonnensystemforschung,
         Katlenburg-Lindau, Germany
         $^2$Department of Mathematics, Imperial College London,
         London, UK
         $^3$National
  Center for Atmospheric Research,\footnote{The National
  Center for Atmospheric Research
  is sponsored by the National Science
  Foundation}
         Boulder, Colorado, USA
         $^4$Departamento de F\'\i sica, Facultad de Ciencias Exactas y Naturales, Universidad de Buenos Aires,
          Ciudad Universitaria, Buenos Aires, Argentina}
%
%


\begin{abstract}

The subfilter-scale (SFS) physics of regularization models are
investigated to understand the regularizations' performance as SFS
models.  Suppression of spectrally local SFS interactions
and conservation of small-scale circulation in the
Lagrangian-averaged Navier-Stokes $\alpha-$model (LANS$-\alpha$) is
found to lead to the formation of rigid bodies.  These contaminate the
superfilter-scale energy spectrum with a scaling that approaches
$k^{+1}$ as the SFS spectra is resolved.  The Clark$-\alpha$ and
Leray$-\alpha$ models, truncations of LANS$-\alpha$, do not conserve
small-scale circulation and do not develop rigid bodies.
LANS$-\alpha$, however, is closest to Navier-Stokes in intermittency
properties. {All three models are found to be stable at high Reynolds
number.  Differences between $L^2$ and $H^1$ norm models are clarified.}
  For magnetohydrodynamics (MHD), the presence of the
Lorentz force as a source (or sink) for circulation and as a
facilitator of both spectrally nonlocal large to small scale
interactions as well as local SFS interactions prevents the formation
of rigid bodies in Lagrangian-averaged MHD (LAMHD$-\alpha$).
LAMHD$-\alpha$ performs well as a predictor of superfilter-scale
energy spectra and of intermittent current sheets at high Reynolds
numbers.  It may prove generally applicable as a MHD-LES.

keywords: LES, Subgrid-scale processes, alpha models, MHD, intermittency
\end{abstract}

\section{Introduction}
\label{SEC:INTRO}

{Computing solutions to the Navier-Stokes equations at realistic
  values of the Reynolds number  ($Re\equiv UL_0/\nu$, with characteristic
velocity, $U$, and length-scale, $L_0$ and
viscosity, $\nu$) for most geophysical, astrophysical,
  and many engineering applications is technologically infeasible.
  This is because the range of dynamically important length (and time)
  scales is quite large: from the large scale, $L_0$, defined by the
  problem down to the scales of viscous dissipation,
  $l_\nu\sim L_0Re^{-3/4}$.  One approach is to simply cut off the
  smallest scales to arrive at a problem small enough for modern
  computational limits.  A low-bandpass filter, $L:
  f\rightarrow\bar{f}$, replaces the velocity, $\vec{v}$, and pressure,
  $P$, with smoother, resolvable fields, $\bar{\vec{v}}\,,\bar{P}$.
  Their time evolution is governed by the filtered Navier-Stokes  equations,
\begin{eqnarray} \partial_t\bar{\vec{v}} + \bar{\boldsymbol{\omega}} \times \bar{\vec{v}}
 =  - \boldsymbol{\nabla} \bar{P} + \nu \nabla^2 \bar{\vec{v}} - \vec{\nabla}\cdot\bar{\tau} \nonumber \\
\nabla\cdot\vec{v}=0\,,
\label{eq:LES}
\end{eqnarray}
where $\boldsymbol{\omega}=\nabla\times\vec{v}$ is the vorticity and $\bar\tau$ is the Reynolds subfilter-scale (SFS) stress tensor,
\begin{equation}
\bar\tau_{ij} = \overline{v_iv_j}-\bar{v}_i\bar{v}_j\,.
\label{eq:tauNS}
\end{equation}
The modeling of the unresolved stress, Eq. (\ref{eq:tauNS}), is the
main challenge of conducting such large eddy simulations (LES, see
\cite{MK00} for a recent review).  }

Regularization modeling {(of the SFS stress tensor)} for Navier-Stokes
{\cite{CFH+98,HMR98b,FHT01,GH03,CHT05,GH06,KiCaAl+2009,LeRaTi2009,RaTi2009},
  magnetohydrodynamics (MHD) \cite{H02a}, Boussinesq convection
  \cite{vaReJoHa2009}, and inviscid cases \cite{LaTi2009}} promises
several advantages.  For Navier-Stokes, only weak, possibly non-unique
solutions have been rigorously proven to exist, \neu{and this can
  impact the possibility of achieving a direct numerical solution
  (DNS), e.g., with Fourier methods \cite{Gu2008}.  This is because to
  prove, even in the linear case, the convergence of a numerical
  solver for a partial differential equation (PDE), one assumes the
  PDE is well-posed (i.e., the existence of unique solutions that
  depend continuously on the data) \cite{LaRi1956}.} A regularization
has strong, unique solutions \neu{(it is well-posed) and the concerns
  of \cite{Gu2008} do not prevent us from achieving} a DNS of the
model equations, \neu{even with Fourier methods.}  It is worth noting
that \neu{numerical convergence of the DNS of the regularization model equations
  implies} a grid-independent model \neu{of the Navier-Stokes
  equations. Additionally, the dissipative term will be unmodified (unlike many
  LES); the Reynolds number remains well defined (e.g.,
  $\boldsymbol{\omega} \times \bar{\vec{v}}/\nu\nabla^2\vec{v}\sim
UL_0/\nu\equiv Re$).  This is opposed to} the usual approach of
  modeling the behavior of the flow in the limit of very large $Re$.
  Thus, the models may be more applicable to intermittent phenomena
  where the length of the inertial range can be important
  \cite{PGHM+06}.  Since the models do not introduce the effect of the
  small scales in an {\sl ad hoc} fashion but rather preserve the
  mathematical properties of the underlying equations, their
  application can further our understanding of turbulence and
  turbulence modeling.  The methods are also more easily generalized
  to other problems (e.g., coupling to a magnetic field).

In this paper we address two separate questions.  One is the question
of the practical applicability of regularization models as SFS models.
When addressing this question, the filter width, $\alpha$, will be
placed in the inertial range and the grid spacing, $\Delta$, will be
just small enough to achieve a DNS of the regularization.  Our aim is
then to determine how well the model's DNS reproduces a ``DNS'' of
Navier-Stokes compared {\sl at scales larger than $\alpha$.}  Our
second question is ``How do the models work?''  \neu{To answer this
  question, we take an approach that is the antithesis of a practical
  LES: we choose the filter width, $\alpha$, to be a large fraction of
  the computation domain (and, thus, a large multiple of $\Delta$).
  Such a calculation is not a LES--there is no superfilter-scale
  inertial range to compare with Navier-Stokes.  Instead, we make a
  new type of study to understand
  the new SFS physics introduced by the regularization.  The
  differences in physics between the SFS model and Navier-Stokes (or
  MHD) is what allows the model to reduce computational cost when it
  is employed as a SFS model. 
  Understanding of how the models work (or fail) can guide
  the development of new models as we will show.}

\section{Navier-Stokes}

\subsection{LANS$-\alpha$ and rigid body formation}
\label{sec:2}

The first model {we consider} is the Lagrangian-averaged Navier-Stokes
(LANS) $\alpha-$model \cite{CFH+98,HMR98b}.  It is derived by Lagrangian averaging fluid motions followed by
application of Taylor's frozen-in turbulence approximation as the
model's one and only closure: fluctuations about the Lagrangian mean
smaller than $\alpha$ are swept along by the large-scale flow and are not
allowed to interact with one another \neu{\cite{H02a}.}  The model is attractive as it
retains the Hamiltonian structure of Navier-Stokes, preserves Kelvin's
theorem (conserves small-scale circulation in the absence of
dissipation), and conserves both total energy and helicity (the correlation between the velocity, $\vec{v}$, and its curl, the vorticity $\boldsymbol{\omega}=\nabla\times\vec{v}$) \neu{\cite{HMR98b}.}  These
properties are conserved in the $H^1_\alpha$ norm instead of the usual
$L^2$ norm.  This is essential when interpreting results of the model
as, for example, quantities involving the square velocity, $|\vec{v}|^2$, must
now be replaced with the dot product $\vec{v}\cdot\bar{\vec{v}}$ where
$\bar{\vec{v}}$ is the {filtered} velocity.  Physically, \neu{due to the frozen-in approximation,} the model
retains \neu{spectrally} nonlocal interactions {(important at finite $Re$ \cite{MiAlPo2008,AlEy2009})} between the superfilter and
subfilter scales while the flux of energy in subfilter scales is
reduced by the limit on local small-scale to small-scale interactions \neu{\cite{FHT01,PGHM+07a}.}

The LANS$-\alpha$ model is given by, {
\begin{eqnarray}
 \partial_t{\vec{v}} + {\boldsymbol{\omega}} \times \bar{\vec{v}}
 =  - \boldsymbol{\nabla} \pi + \nu \nabla^2 {\vec{v}}  \nonumber \\
\nabla\cdot\vec{v}=\nabla\cdot\bar{\vec{v}}=0\,.
\label{eq:lans}
  \end{eqnarray}
From the identity, ${\boldsymbol{\omega}} \times \bar{\vec{v}}=
\bar{\vec{v}}\cdot\nabla\vec{v}+(\nabla\bar{\vec{v}})^T\cdot\vec{v}
-\nabla(\bar{\vec{v}}\cdot\vec{v})$,
we see that it}
differs from Navier-Stokes both in advection by the smoothed velocity
and the addition of a second nonlinear term associated with the conservation
of circulation.  Traditionally, LANS$-\alpha$ is used with an inverse Helmholtz
operator as the filter: $\bar{v}_i = (1 - \alpha^2\partial_{jj})^{-1}v_i$.
In this case, LANS$-\alpha$ can be written as a LES, {Eq. (\ref{eq:LES}),} with
\begin{equation}
\bar{\tau}_{ij}^\alpha = (1 - \alpha^2\partial_{jj})^{-1}\alpha^2
        (\partial_m\bar{v}_i\partial_m\bar{v}_j+\partial_m\bar{v}_i\partial_j\bar{v}_m-\partial_i\bar{v}_m\partial_j\bar{v}_m)\,.
\label{eq:sfs}
\end{equation}
 The model allows for
a reduction in resolution without changing \neu{(or supplementing)} the dissipative
term by \neu{instead} altering the SFS scaling properties.  Near {wavenumber,}
{$k=2\pi/\alpha$,} the {$H_\alpha^1$ energy} spectrum is predicted to transition from $k^\beta$
with $\beta=-5/3$ at larger scales to $\beta={-1}$ at smaller scales
\cite{FHT01}.  Consequently, dissipation \neu{($\Omega_{(\alpha)}(k)=k^2E_{(\alpha)}(k)$)} goes as $k^1$ instead of
$k^{1/3}$ and the same amount of power is dissipated in fewer scales.
The change in spectral scaling also predicts a resolution requirement in degrees
of freedom, $dof$, for LANS \cite{FHT01},
\begin{equation}
dof_\alpha\sim\alpha^{-1}Re^{3/2}\,,
\end{equation}
which has been confirmed in numerical experiments \cite{PGHM+07a}.
\neu{Once $dof_\alpha$ has been resolved, further resolution yields no change in the numerical solution: LANS$-\alpha$ is} a grid-independent
SFS model.
When compared with the $dof$ for Navier-Stokes,
\begin{equation}
dof_{NS}\sim Re^{9/4}\,,
\end{equation}
we see that LANS$-\alpha$ should improve as a SFS model for larger $Re$.
This was an encouraging prediction as {LANS$-\alpha$} compared well with
dynamic eddy viscosity \cite{MKS+03}
and dynamic mixed (similarity) eddy viscosity \cite{GH06} at moderate $Re$.

\begin{figure}[htbp]
\centering
    \includegraphics[width=0.7125\linewidth]{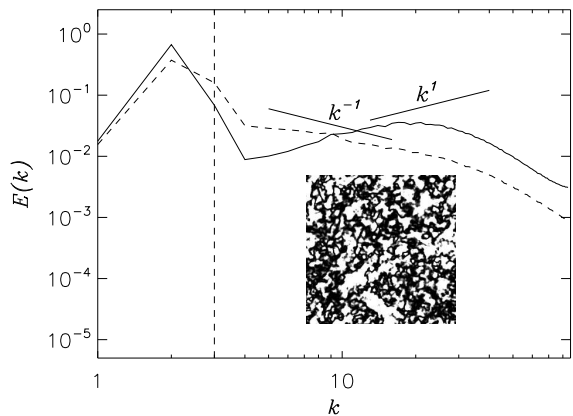}
\caption{Energy spectrum for LANS$-\alpha$ (solid line) with filter
  scale $\alpha=2\pi/3$ (vertical dashed line).  The SFS power law is well approximated by $k^{+1}$.  Insert shows
  thresholded cubed velocity increment $|\delta\bar{v}_\|(2\pi/10)|^3<10^{-2}$ in
  black.  These regions do not contribute to the turbulent cascade of
  energy to smaller scales and are identified with rigid bodies.  A
  spectrum of only the white regions (dashed line) is consistent with
  the predicted $k^{-1}$ scaling outside rigid bodies.}
\label{fig:nugget}
\end{figure}

We have, however, found that LANS$-\alpha$ develops a problem at large
$Re$: it develops a positive-exponent power-law bump in its
small-scale energy spectrum and a contamination of superfilter-scale
spectral properties \cite{PGHM+07a}.  To investigate
the SFS physics responsible for this, we employ \neu{(as the antithesis of a LES)} a filter $1/3$ the
size of our $256^3$ computational cube in a pseudo-spectral
calculation \cite{GMD05b,GMD05} with a Taylor-Green (TG) forcing \cite{TG37} and
$Re\approx8000$.  As shown in Fig. \ref{fig:nugget}, the observed
scaling law is $k^{+1}$.  This was shown to be associated with the formation {in the flow of passively swept regions, called \emph{rigid bodies} \cite{PGHM+07a}.}  These form as a consequence of disallowing
sub-$\alpha$-scale fluctuations to interact with each other in the
closure approximation.  A rigid body cannot support longitudinal
velocity increments:
$\delta\bar{v}_\|(l)\equiv[\bar{\vec{v}}(\vec{x})-\bar{\vec{v}}(\vec{x}+\vec{l})]\cdot\vec{l}/l=0$.
This \neu{dimensionally predicts a scaling relation, $\delta\bar{v}\sim l^0$,
and, with $v\sim(1+\alpha^2k^2)\bar{v}\sim\alpha^2k^2\bar{v}$} for $l\ll\alpha$, an energy
spectrum of
\begin{equation}
E_\alpha(k) \sim \bar{v}vk^{-1} \sim k^1
\end{equation}
which is compatible with the observed SFS energy spectrum.  Inside
rigid bodies there can be no turbulent cascade of energy to smaller
scales (\neu{since there are} no internal degrees of freedom).  From the K\'arm\'an-Howarth
theorem, we should then expect to be able to detect rigid bodies by
visualizing the cubed velocity increments (which are proportional to
the energy flux).  The regions which correspond to negligible flux are
shown as black in the inset of Fig. \ref{fig:nugget}.  Filtering these
regions out, allows us to obtain a (convolved) energy spectra for the
remaining white portion of the flow.  This spectrum is shown as a
dashed line in Fig. \ref{fig:nugget} and has a negative spectral slope
close to the predicted $k^{-1}$ spectrum.  The
resulting picture of the model's behavior is to produce two spatially
separate scalings. The white portions of the flow possess the
predicted LANS$-\alpha$ scaling and are responsible for the
observation of the predicted $dof_\alpha$.  The black portions are
rigid bodies whose $k^1$ energy spectrum dominates over $k^{-1}$ for
large $k$ and are responsible for the observed spectral contamination.
\neu{Because of this spectral contamination, there is no need for further tests of LANS$-\alpha$ against newer models such as the Variational Multiscale Method \cite{HuMaOb+2001}.  Note, however,
that suitable spectra can be obtained with a modified viscous length scale (LANS$-\alpha\beta$
  \cite{KiCaAl+2009}).}

\subsection{The influence of circulation on rigid bodies}

\begin{figure}[htbp]
\centering
    \includegraphics[width=0.7125\linewidth]{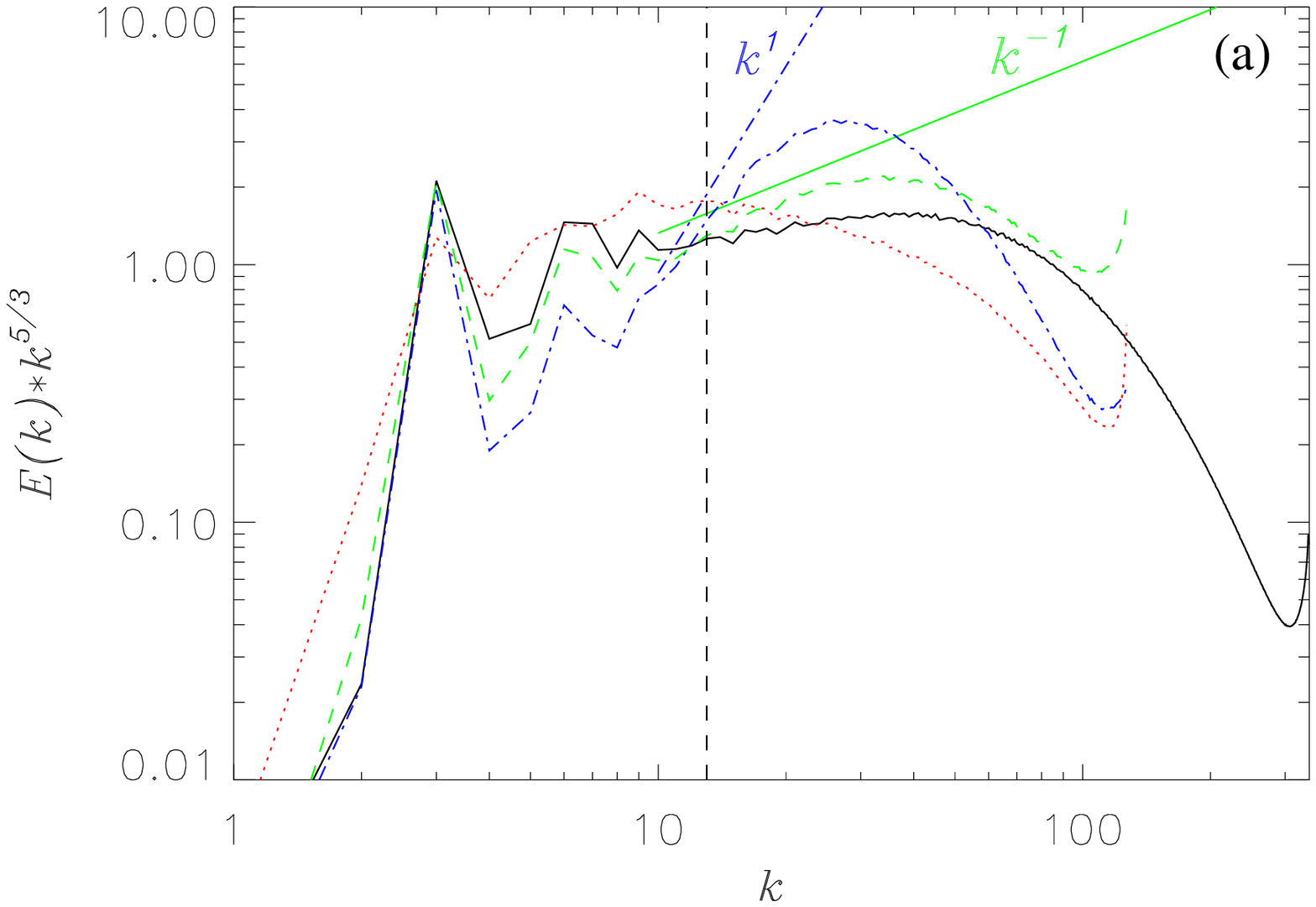}
    \includegraphics[width=0.7125\linewidth]{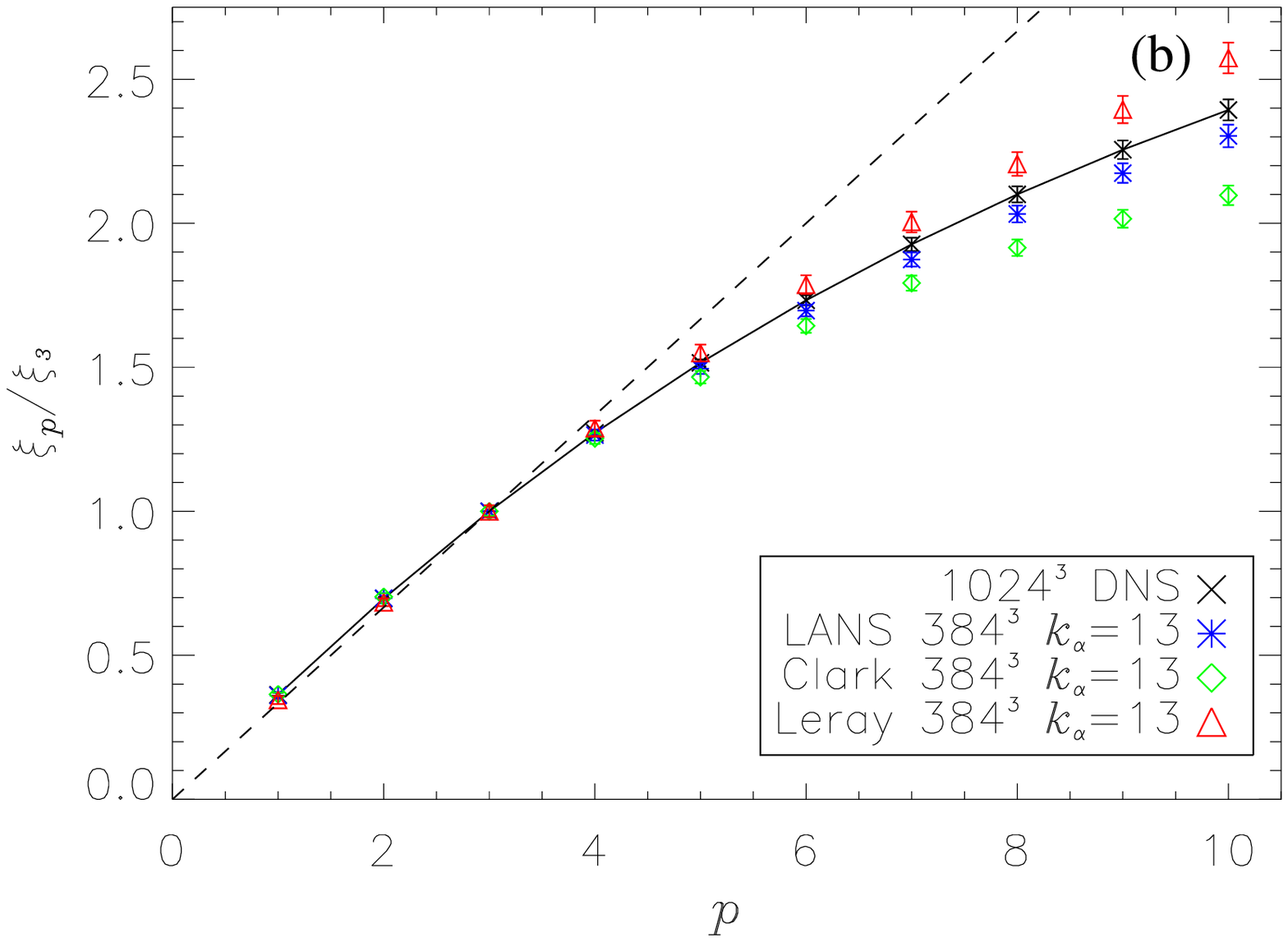}
\caption{{(a)} Compensated energy spectra {($2\pi/\alpha$, vertical dashed
    line)} for Navier-Stokes (solid black), LANS$-\alpha$
  (blue dash-dotted), Clark$-\alpha$ (green dashed), and Leray$-\alpha$
  (red dotted).  {(b)} Normalized structure function scaling exponent
  $\xi_p/\xi_3$ versus order $p$.  Clark$-\alpha$ is the
  best approximation for the superfilter-scale
  spectrum, whereas {high-order} intermittency properties
  are best reproduced by LANS$-\alpha$ \cite{PiGrHoMi+2008}.}
\label{fig2}
\end{figure}

The formation of rigid bodies in LANS$-\alpha$ limits the reduction of
numerical $dof$ saved compared to Navier-Stokes to a factor of $1/12$
regardless of $Re$ \cite{PGHM+07a}.  It is desirable, then, to
alter the model in such a way to prevent the formation of rigid
bodies.  Truncation of the SFS stress tensor,
Eq. (\ref{eq:sfs}), to the first term results in the Clark$-\alpha$
model \cite{CHT05}, {
\begin{eqnarray}
 \partial_t{\vec{v}} + (1-\frac{1}{2}\alpha^2\nabla^2)({\bar{\boldsymbol{\omega}}} \times \bar{\vec{v}})-\frac{1}{2}\alpha^2\left[
(\nabla^2{\bar{\boldsymbol{\omega}}})\times\bar{\vec{v}}+
\bar{\boldsymbol{\omega}}\times(\nabla^2{\bar{\vec{v}}})\right]
 =  \nonumber \\  - \boldsymbol{\nabla} \mathcal{P} + \nu \nabla^2 {\vec{v}}  \nonumber \\
\nabla\cdot\vec{v}=\nabla\cdot\bar{\vec{v}}=0\,,
\label{eq:clark}
  \end{eqnarray}
} and to the first two terms results in the
Leray$-\alpha$ model \cite{GH03,GH06}, {
\begin{eqnarray}
 \partial_t{\vec{v}} + \bar{\vec{v}}\cdot\nabla{\vec{v}} 
 =  - \boldsymbol{\nabla} p + \nu \nabla^2 {\vec{v}}  \nonumber \\
\nabla\cdot\vec{v}=\nabla\cdot\bar{\vec{v}}=0\,.
\label{eq:leray}
  \end{eqnarray}
}  Both these models are regularizations and conserve the
total energy of the flow \neu{\cite{CHT05,GH03,GH06}.}  They do not, however, conserve the helicity
nor the small-scale circulation.  {Considering the rotational properties
of a rigid body (in the absence of viscous friction),} these models' circulation
properties may be incompatible with rigid body formation.
\neu{Indeed,
while LANS$-\alpha$ exhibits a positive-exponent
power law in this case ($\alpha=2\pi/13$, $Re\approx3300$, TG forcing), both Clark$-\alpha$ and Leray$-\alpha$ are free
from this signature of rigid body formation (Fig. \ref{fig2}a).}  However, LANS$-\alpha$'s
intermittency properties are more similar to Navier-Stokes than
the other two models (Fig. \ref{fig2}b).

\subsection{Computability and interpretation of $H_\alpha^1$ norm regularizations}

\begin{figure}[htbp]
\centering
  \includegraphics[width=0.7125\linewidth]{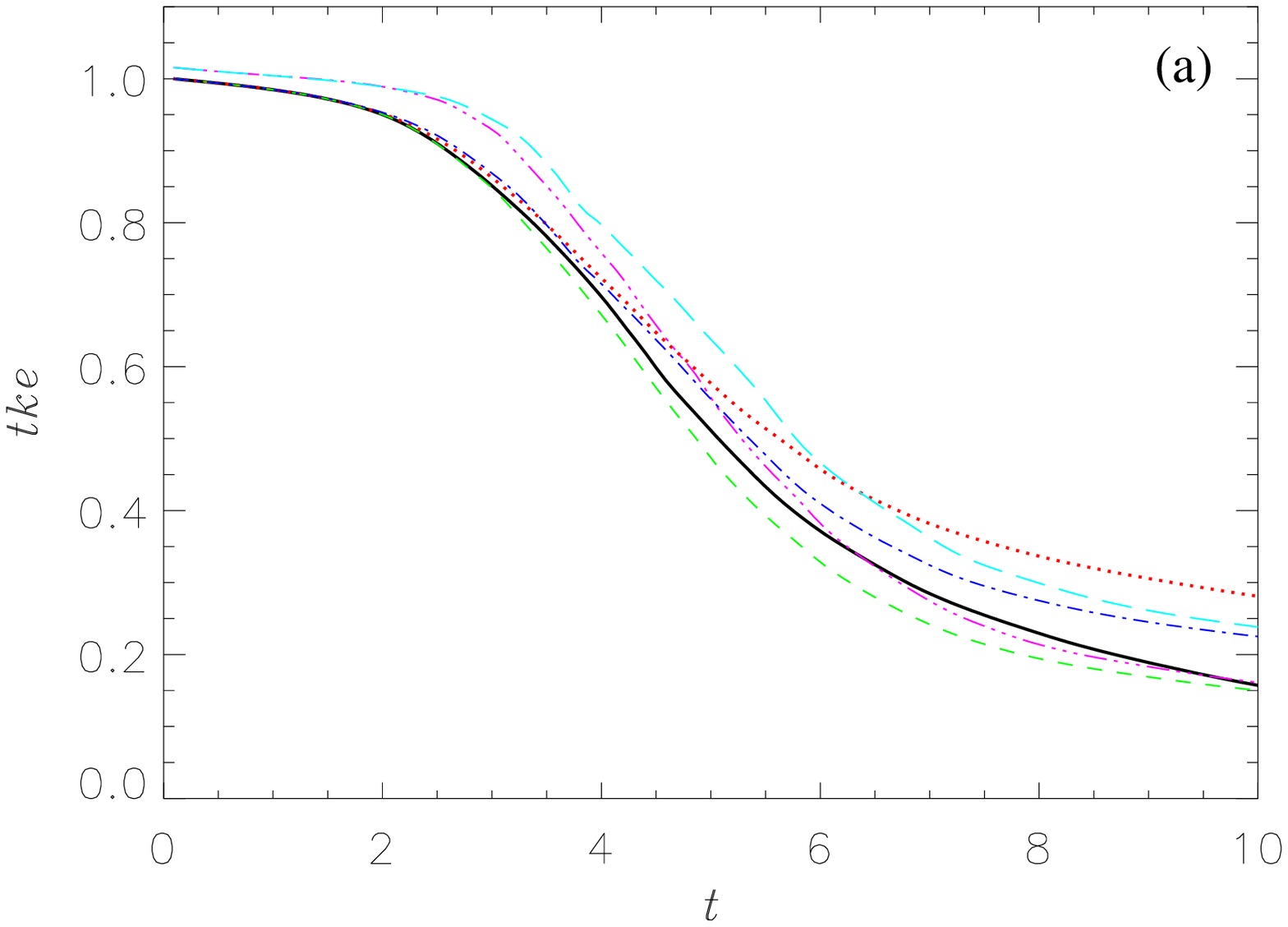}
  \includegraphics[width=0.7125\linewidth]{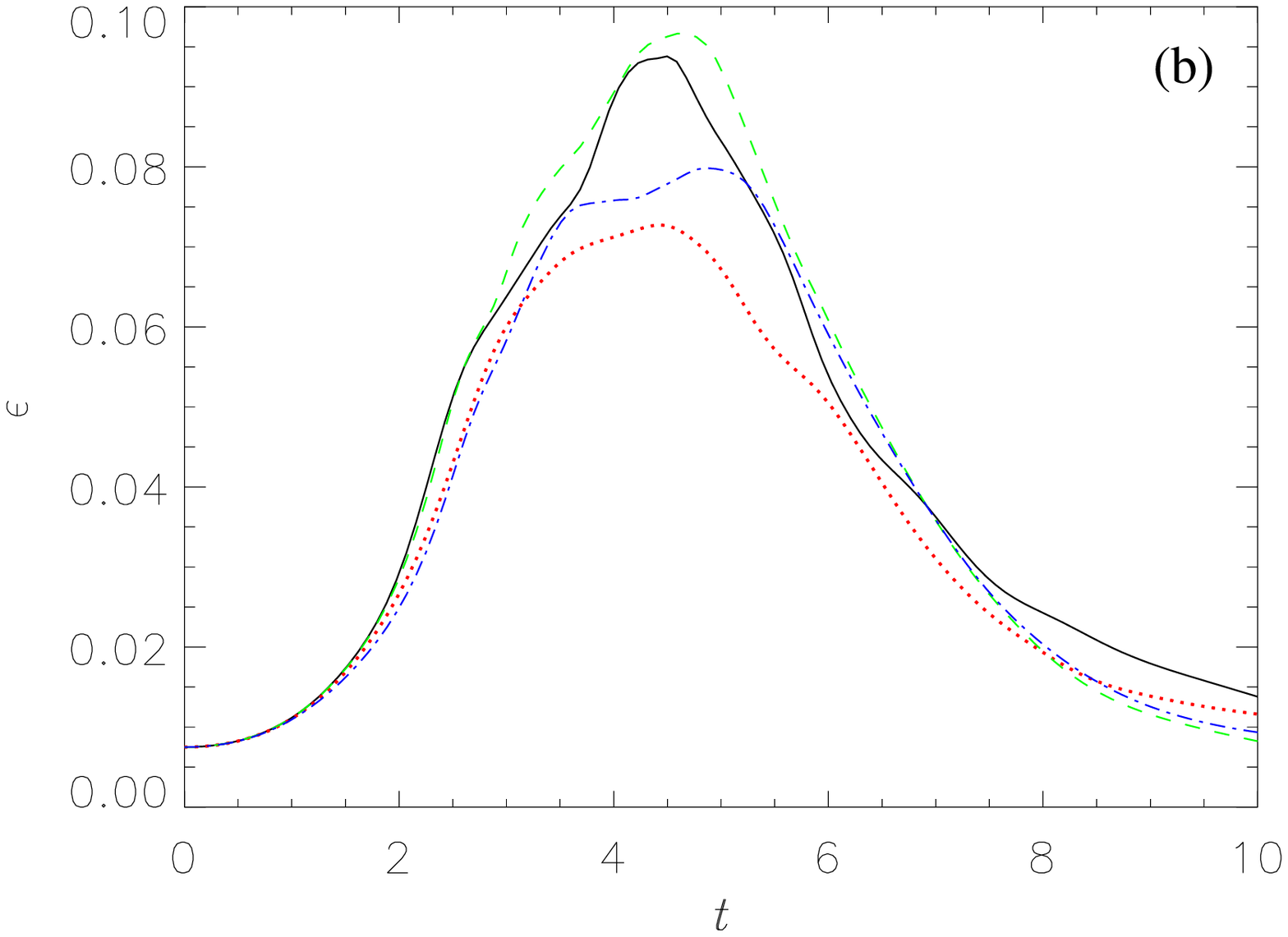}
  \caption{Case3a, $Re\approx400$.  Line styles are as in
    Fig. \ref{fig2}a for correct norms with the addition of $L^2$ norm
    energy for Clark$-\alpha$ (pink dash-triple-dotted) and for
    LANS$-\alpha$ (cyan long-dashed). (a)
    Filtered kinetic energy, $tke$.  (b) Dissipation,
    $\varepsilon=dE/dt$, versus time, $t$.}
  \label{fig:chfr_vs_tP1}
\end{figure}

\begin{figure}[htbp]
\centering
  \includegraphics[width=0.7125\linewidth]{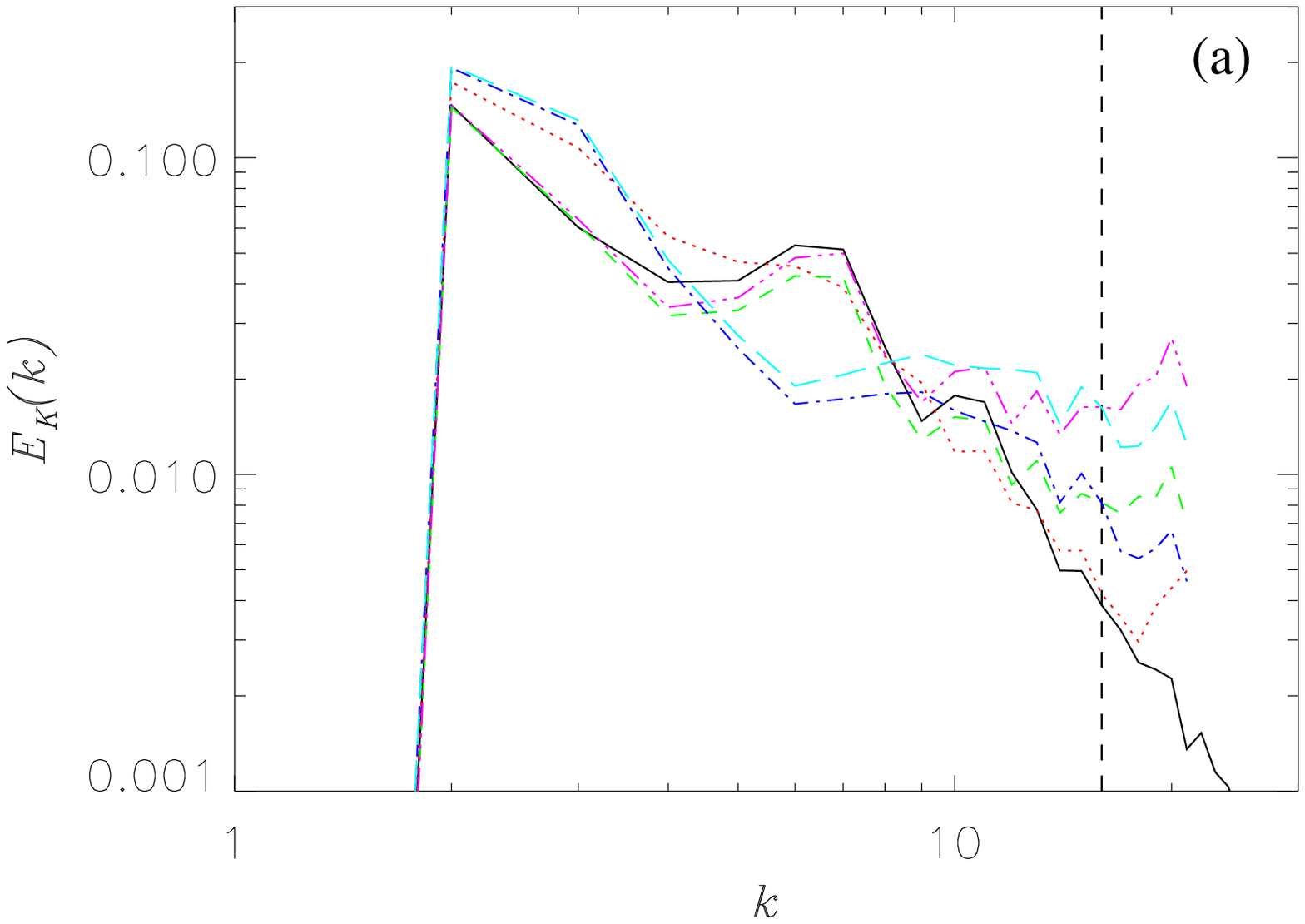}
  \includegraphics[width=0.7125\linewidth]{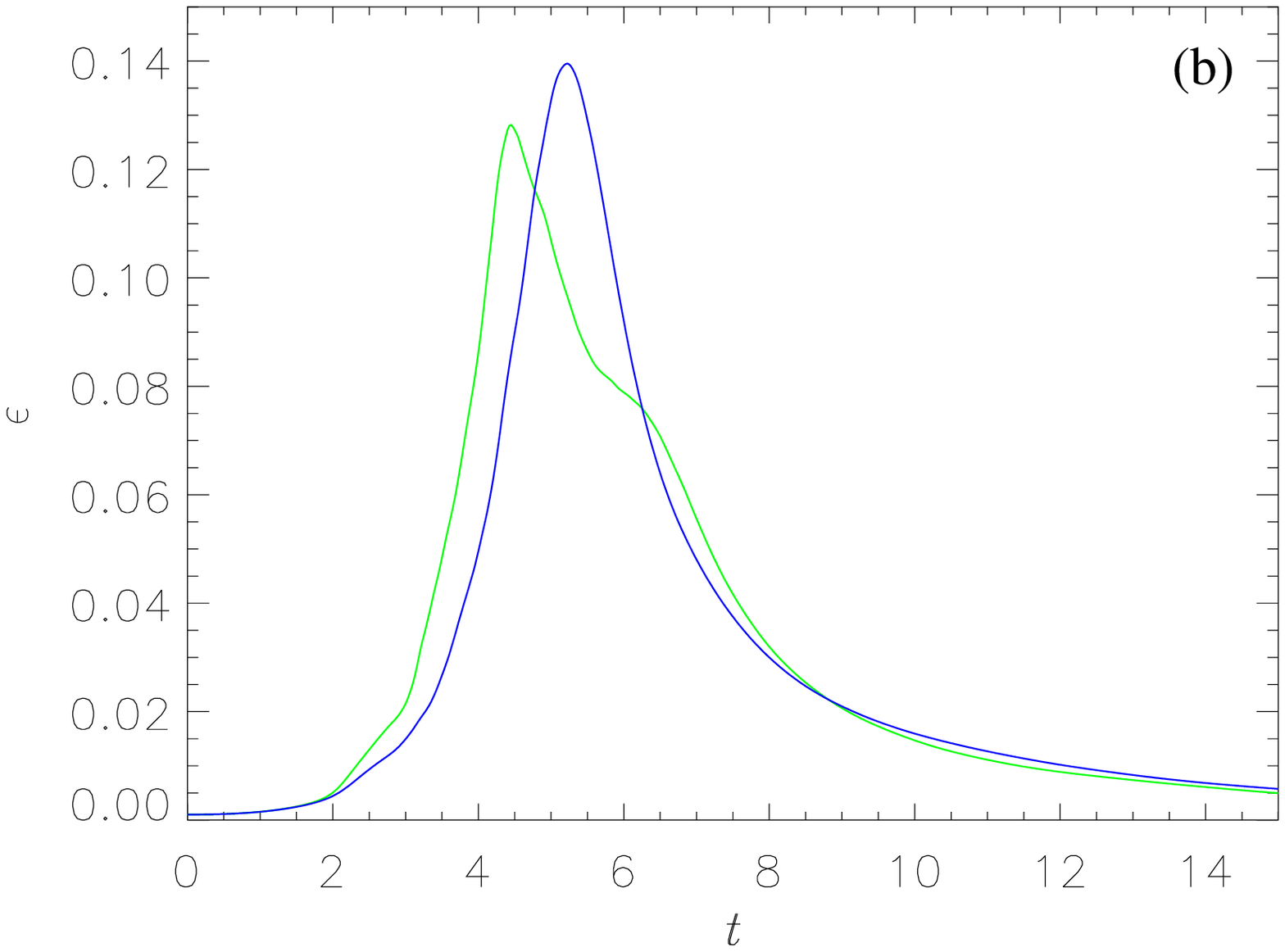}
  \caption{(a) Case3a, $Re\approx400$: energy spectra after peak of
    dissipation, $t\in[4.5,5.5]$.  Line styles are as in
    Fig. \ref{fig:chfr_vs_tP1}.  Vertical dashed line is
    $2\pi/\alpha$.  (b) Case3b, $Re\approx3000$: dissipation,
    $\varepsilon$, versus time, $t$ for Clark$-\alpha$ (light green)
    and for LANS$-\alpha$ (dark blue). }
  \label{fig:chfr_vs_tP2}
\end{figure}

It has been reported recently that Clark-$\alpha$ and LANS-$\alpha$
have poor SFS model performance and, in fact, have instabilities
\neu{in simulations with periodic boundary conditions
  \cite{ChFr2009}.} This result {highlights the importance of
  correctly interpreting $H_\alpha^1$ norm regularizations.  To
  illustrate this, we compute $256^3$ DNS and $64^3$ models runs for
  their case 3a: decaying Taylor-Green vortex with
  $\nu=2.5\cdot10^{-3}$ (time step, $dt=3\cdot10^{-3}$ for all
  simulations).  In Figs. \ref{fig:chfr_vs_tP1} and
  \ref{fig:chfr_vs_tP2}, we display results for the models employing a
  filter width $\alpha=2\pi/16$ (results with $\alpha=2\pi/24$ are
  closer to that of the DNS).  Fig. \ref{fig:chfr_vs_tP1}a shows the
  time evolution of the total kinetic energy in the super-filter
  scales,
  \begin{equation}
    tke \equiv \int_0^{k_\alpha} E_{(\alpha)}(k)\,,
  \end{equation}
  where $k_\alpha=2\pi/\alpha$.  Incorrectly applying the $L^2$ norm,
  $E=|\vec{v}|^2/2$, to Clark$-\alpha$ and LANS$-\alpha$ leads to the
  interpretation that they deviate significantly from the DNS due to
  slower energy decay.  Correctly applying the $H_\alpha^1$ norm,
  $E_\alpha=\bar{\vec{v}}\cdot\vec{v}/2$, shows that the models are,
  in fact, doing quite well as can be seen in the plot of the
  dissipation rate of energy, $\varepsilon$
  (Fig. \ref{fig:chfr_vs_tP1}b).  The qualitative properties of
  Lagrangian models is that they have the same invariants as the
  primitive equations, albeit in a different norm. Since it is these
  invariants which may very well influence the dynamics, as for
  example shown in \cite{CiBoDe+2005} where Kolmogorov spectra are
  present at intermediate times at large scale in ideal (Euler)
  three-dimensional fluid flows, one must compare the behavior of one
  set of invariants to the other set in the Lagrangian modeling
  formulation.  Similarly, the $L^2$ norm spectra for Clark$-\alpha$
  and LANS$-\alpha$ in Fig. \ref{fig:chfr_vs_tP2}a appear
  under-dissipative at high wave numbers, while the correct
  $H_\alpha^1$ norm spectra are closer to the DNS. We also conduct
  Clark$-\alpha$ and LANS$-\alpha$ computations for case 3b of
  \cite{ChFr2009}: $\nu=1/3000$, $\alpha=2\pi/32$ and $384^3$ grid
  points \neu{(Fig. \ref{fig:chfr_vs_tP2}b).}  We find no signs of
  instability.  The particular numerical expression of the models we
  used is given in Eqs. (\ref{eq:lans}) and (\ref{eq:clark}).  These
  differ from Eqs. (8) and (9) in \cite{ChFr2009} and it is known that
  discretized pseudospectral operators depend on their algorithmic
  form (e.g., $\nabla\psi^2\neq2\psi\nabla\psi$
  \cite{DaMoDo+1986,CHQ+88}).  This is likely the source of their
  observed instability; however a recently discovered deficiency in
  high-order low-storage Runge-Kutta schemes \cite{BrMiRo+2008} may
  also impact numerical implementation of the models.  For this
  reason, our calculations are made with second-order Runge-Kutta in
  time.}  \neu{It is worth pointing out that Clark$-\alpha$ is nearly
  identical to the Rational Large Eddy Simulation (RLES) model
  \cite{GaLa2000} which has been shown to be unstable for non-periodic
  boundaries \cite{Berselli2004,John2005}.  Both our work and that of
  \cite{ChFr2009} have employed periodic boundary conditions only, but
  extra precautions must be taken for stability in the non-periodic
  case \cite{Berselli2004,John2005}.}

\section{MHD: circulation and outlook for LES}

{In MHD, the \neu{circulation properties are} quite different since small-scale
  circulation is broken by the Lorentz force $\vec{j} \times \vec{b}$,
  with $\vec{j}=\boldsymbol{\nabla} \times \vec{b}$ the current,
  $\vec{b}$ being the induction. This force acts as source
  (sink) of circulation, $\Gamma$, as opposed to the insufficient modeling of $\Gamma$ in the Leray$-\alpha$
  and Clark$-\alpha$ models. This can be seen in Kelvin's theorem,}
\begin{equation}
\frac{d}{dt} \Gamma = \frac{d}{dt} \oint_{\cal C} \vec{v}\cdot
d\vec{r} = \oint_{\cal C} \vec{j} \times {\vec{b}} \cdot d\vec{r}\,.
\end{equation}
This may prevent the formation of rigid bodies even while conserving
all the correct physical properties of the flow.

The LES equations for MHD are given by
\begin{eqnarray} \partial_t\bar{\vec{v}} + \bar{\boldsymbol{\omega}} \times \bar{\vec{v}}
 = \bar{\vec{j}} \times \bar{\vec{b}} - \boldsymbol{\nabla} \bar{\Pi} + \nu \nabla^2 \bar{\vec{v}} - \vec{\nabla}\cdot\bar{\tau} \nonumber \\ 
\partial_t \bar{\vec{b}} = \boldsymbol{\nabla} \times \left( \bar{\vec{v}} \times \bar{\vec{b}} \right) + \eta \nabla^2 \bar{\vec{b}} - \vec{\nabla}\cdot\bar{\tau}^b\,,
\label{eq:lamhdLES}
\end{eqnarray}
{where $\eta$ is the magnetic diffusivity,} $\Pi=P+|\vec{b}|^2/2$ the
modified pressure, $\bar\tau$ is the Reynolds SFS stress tensor,
\begin{equation}
\bar\tau_{ij} = \overline{v_iv_j}-\bar{v}_i\bar{v}_j- (\overline{b_ib_j}-\bar{b}_i\bar{b}_j)\,,
\label{eq:tau}
\end{equation}
and $\bar\tau^b$ is the electromotive-force SFS stress tensor,
\begin{equation}
\bar\tau_{ij}^b = \overline{b_iv_j}-\bar{b}_i\bar{v}_j- (\overline{v_ib_j}-\bar{v}_i\bar{b}_j)\,.
\label{eq:taub}
\end{equation}
Note that the extension of eddy viscosity to eddy resistivity employs the usual
form for $\bar\tau$ involving only the filtered velocity while the
expression for $\bar\tau^b$ similarly only involves the filtered magnetic field
\cite{TFS94}.  Meanwhile,
Eqs. (\ref{eq:tau}) and (\ref{eq:taub}) make it explicitly clear that
interactions between the two fields {\sl at subfilter scales} must be
taken into account.

Another problem with
extending  eddy-viscosity concepts to MHD is that they  can be related
to a known power law of the energy spectrum \cite{ChLe1981}.  This is
inappropriate for MHD as neither kinetic nor magnetic energy is a
conserved quantity and the general expression of the energy spectrum
is not known at this time \cite{I64,K65,GoSr1995,MaCaBo2008,LeBrPo+2009}. 
Additionally, MHD has
been shown to have  nonlocal
interactions between large and small (superfilter and subfilter)
scales \cite{AMP05a} {(e.g., in the Batchelor viscous-inductive regime, $\nu\gg\eta$, where energy is transferred directly from viscous-scale, $l_\nu$, eddies to 
 small-scale magnetic field, $l\ll l_\nu$ \cite{AlEy2010}).}  This complex interaction is a challenge in general for MHD-LES, but may
be an advantage for the Lagrangian-averaged approach as energy
exchanges with sub-$\alpha$ scales may disrupt rigid body formation.
 Some limited case {MHD LES}
include the cross-helicity model \cite{MC02} assuming
alignment between the fields and the low magnetic $Re$ LES \cite{KM04,PPP04}.
\neu{Extensions of spectral models to MHD based on two-point closure formulations of the dynamical equations proposed recently look promising in the analysis of turbulent flows and of the dynamo mechanism \cite{BaPoPo+2008}.
Approximate deconvolution models for MHD \cite{LaTr2010}
also appear promising, but have yet to be tested on non-laminar flows.}
However, there are many regimes of MHD dependent on the ratios
between the various conserved quantities {and $\nu/\eta$:}
\neu{there has yet to be demonstrated a}
 generally applicable LES for MHD. 

\subsection{LAMHD$-\alpha$ and absence of rigid bodies}

\begin{figure}[htbp]
\centering
  \includegraphics[width=0.7125\linewidth]{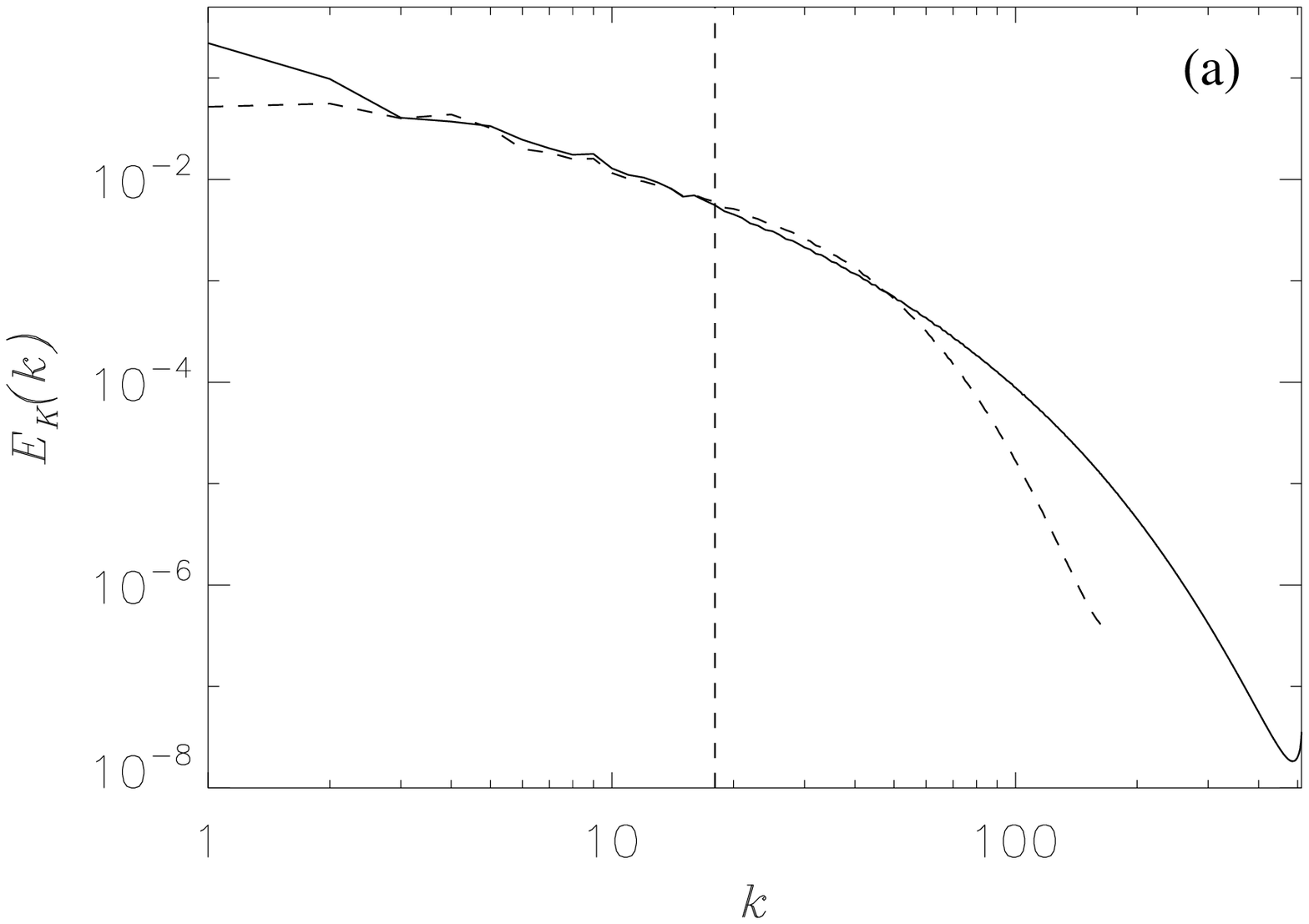}
  \includegraphics[width=0.7125\linewidth]{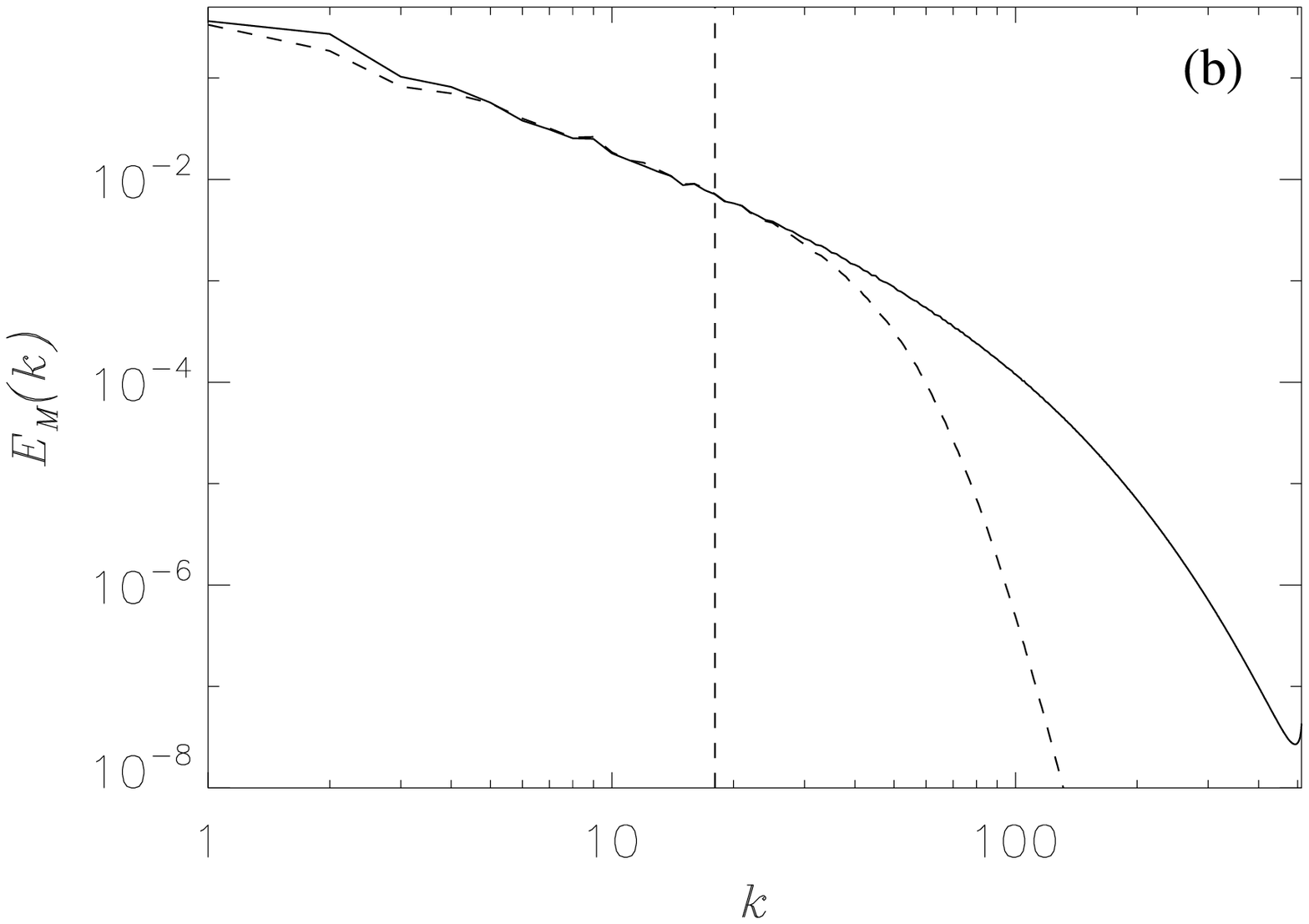} 
  \caption{ Kinetic {(a)} and magnetic {(b)} energy spectra
    {($2\pi/\alpha$ vertical dashed
    line).} The largest scales are affected by differences from the MHD DNS (solid lines) in
    initial conditions. 
    LAMHD$-\alpha$ (dashed lines) exhibits neither the positive power-law nor the
    superfilter-scale spectral contamination associated with high
    $Re$ LANS$-\alpha$.  }
  \label{FIG:COMP1}
\end{figure}

The Lagrangian-averaged MHD $\alpha-$model
(LAMHD$-\alpha$) \cite{H02a,H02b,MP02} is given by, {where the velocity if filtered as before and $\bar{\vec{b}} = (1 -
\alpha_M^2\nabla^2)^{-1}\vec{b}$:}
\begin{eqnarray}
\partial_t\vec{v} + \boldsymbol{\omega}\times\bar{\vec{v}} =
\vec{j}\times\bar{\vec{b}} - \boldsymbol{\nabla}\pi + \nu
\nabla^2\vec{v}\nonumber\\
\partial_t\bar{\vec{b}}
=\boldsymbol{\nabla}\times(\bar{\vec{v}}\times\bar{\vec{b}}) +\eta
\nabla^2\vec{b}\nonumber\\
\boldsymbol{\nabla}\cdot\vec{v}=\boldsymbol{\nabla}\cdot\bar{\vec{v}}=
\boldsymbol{\nabla}\cdot\vec{b}=\boldsymbol{\nabla}\cdot\bar{\vec{b}}=
0\,.
\end{eqnarray}
\neu{LAMHD$-\alpha$ may be written as a MHD-LES, Eqs. (\ref{eq:lamhdLES}), for the case $\alpha_M=\alpha$ \cite{PGHM+06,PiGrMiPo2009}, which we study here.}
The model preserves the ideal quadratic
invariants of MHD (in the $H^1_\alpha$ norm) as well as Alfv\'en's
theorem for frozen-in field lines \neu{\cite{H02a}.}  Physically, it supports Alfv\'en
waves at all scales while slowing and hyper-diffusively damping waves
with wavelengths, $\lambda$, smaller than $\alpha$  \cite{PiGrMiPo2009}.  In
examinations of its SFS physical properties LAMHD$-\alpha$ (dashed lines) displays
neither positive-exponent power-law scaling nor superfilter-scale
spectral contamination (see Fig. \ref{FIG:COMP1}).  Under similar
conditions LANS$-\alpha$ (not shown) displays these signs of rigid body
formation.  Further examinations with larger filters and higher $Re$
were unable to {unravel} any sign that rigid bodies form for
LAMHD$-\alpha$ \cite{PiGrMiPo2009}.

\subsection{LAMHD$-\alpha$ as a SFS model}

\begin{figure}[htbp]
\centering
  \includegraphics[width=0.7125\linewidth]{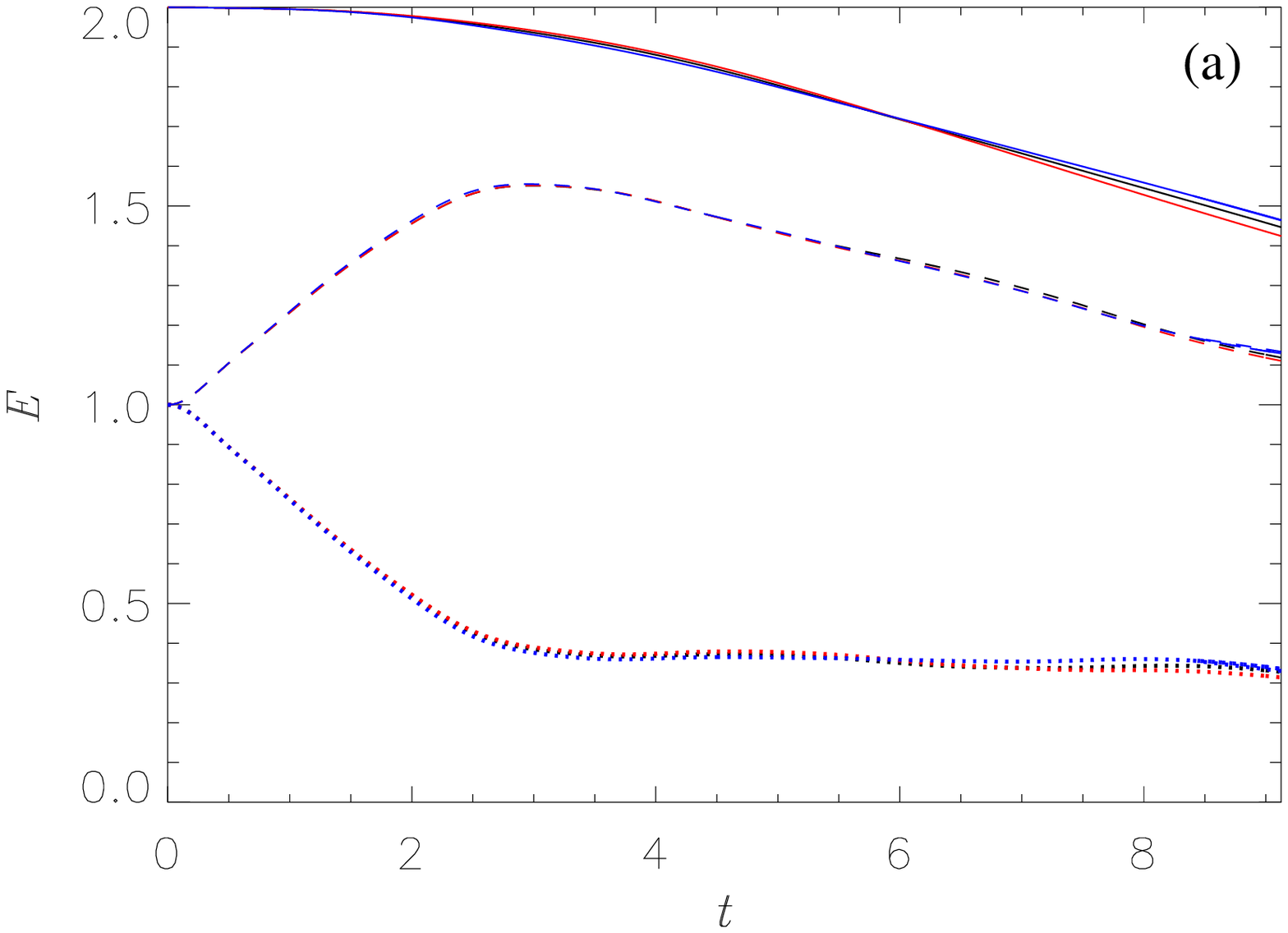}
  \includegraphics[width=0.7125\linewidth]{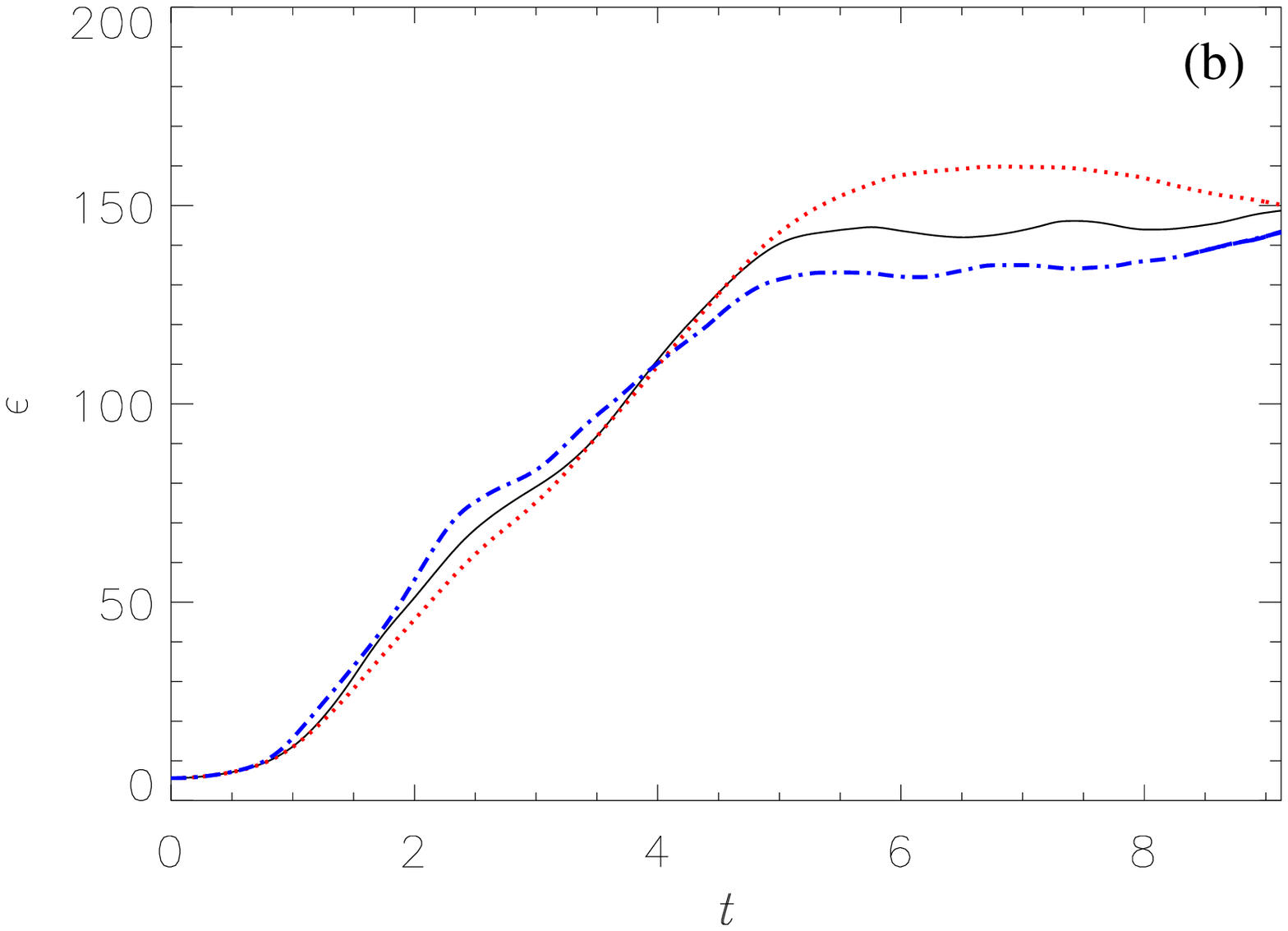} 
  \caption{Time evolution of {total (solid), magnetic (dashed), and kinetic (dotted) energies (a) and total {dissipation, $\varepsilon,$}
    (b) for $1024^3$ DNS (black/solid), $168^3$ LAMHD$-\alpha$
    (blue/dash-dotted), and $168^3$ no-model (red/dotted).}
    LAMHD$-\alpha$ provides no improvement in prediction of global
    quantities over an under-resolved DNS. }
  \label{fig:sfs_global}
\end{figure}

\begin{figure}[htbp]
\centering
  \includegraphics[width=0.7125\linewidth]{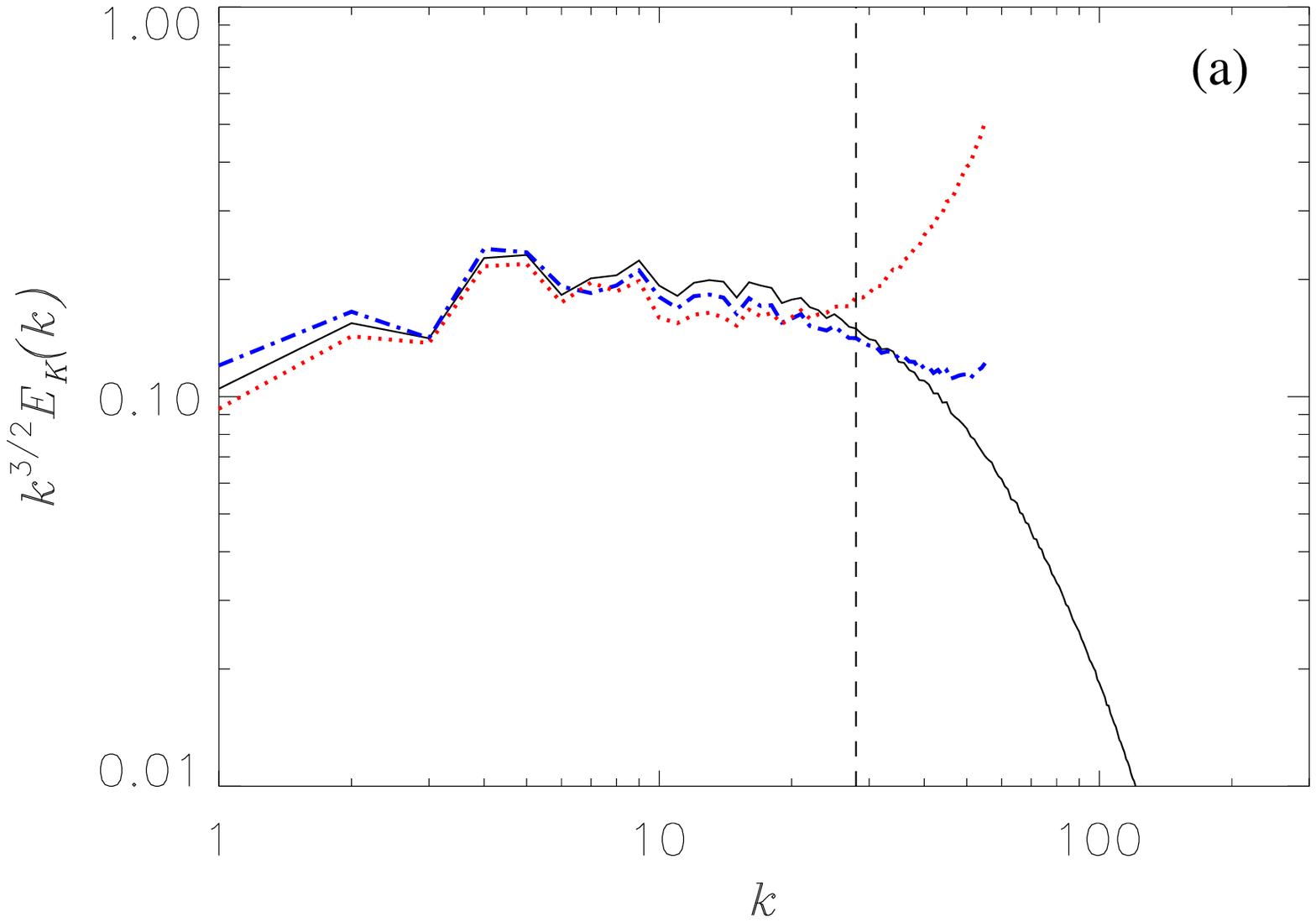}
  \includegraphics[width=0.7125\linewidth]{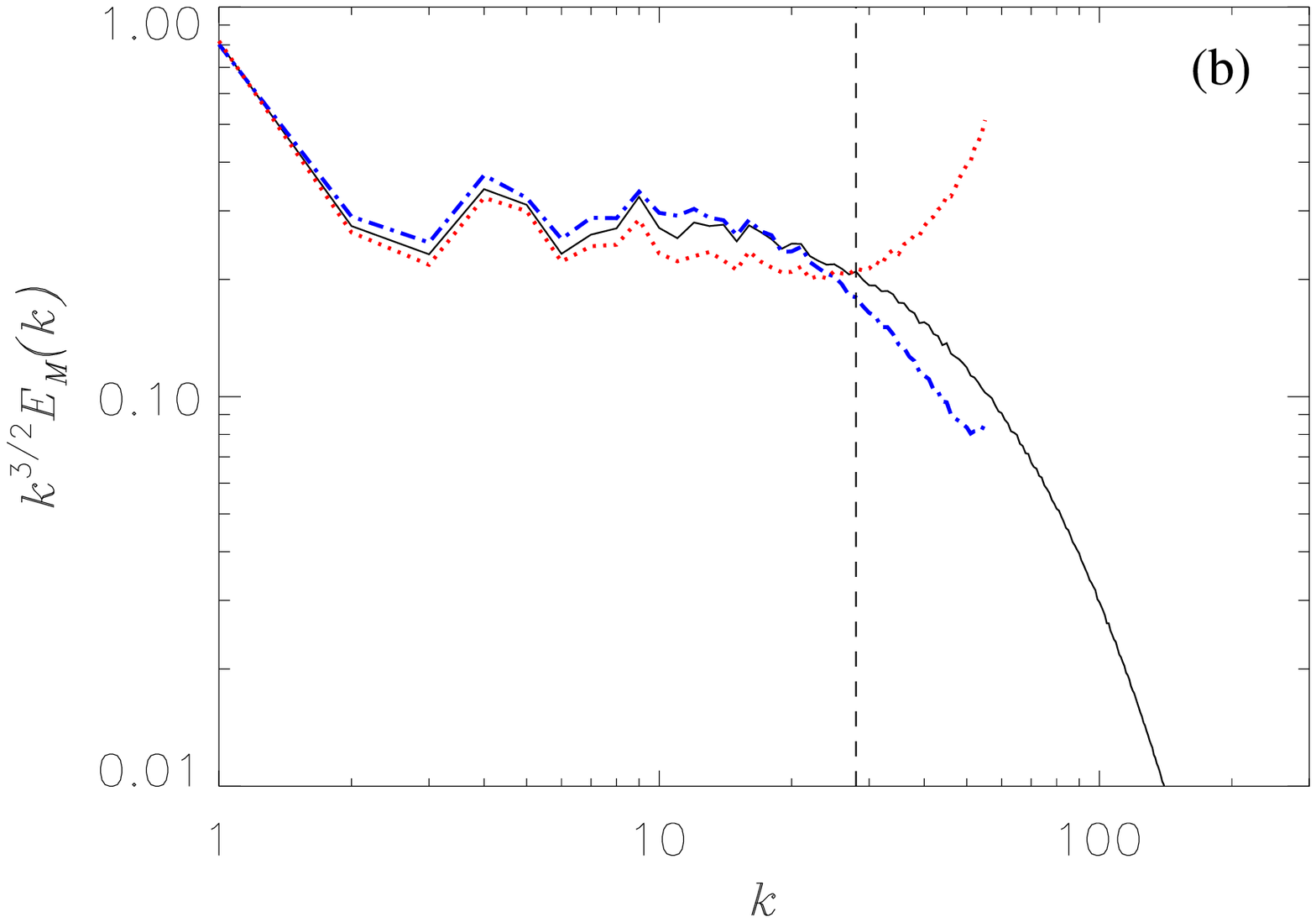} 
  \caption{Compensated kinetic {(a)} and magnetic {(b)} energy
    spectra averaged over $t\in[8,8.8]$.  {Line styles} are as in Fig. \ref{fig:sfs_global}. \neu{($2\pi/\alpha$ vertical dashed
    line).}
    Energy piles up at small scales in the no-model approach {(under-resolved DNS)} and
    LAMHD$-\alpha$ is seen to provide improved predictions of the
    superfilter-scale spectra, especially for the magnetic field.}
  \label{fig:sfs_spectra}
\end{figure}

\neu{Since LAMHD$-\alpha$ did not
display any signs of rigid body spectral contamination,} we test it as a SFS model for large kinetic and
magnetic Reynolds numbers, $\approx3300$.  A DNS of MHD is computed at
a resolution of $1024^3$.  The initial conditions for $\vec{v}$ and
$\vec{b}$ are a superposition of ABC modes \cite{GaFr1986} with
random phases and wavenumbers $k\in[1,4]$.  No external forcing is
applied and the total energy is allowed to freely decay.
LAMHD$-\alpha$ is computed for identical conditions at a resolution of
$N^3=168^3$ with the same $Re$ and a filter size $\alpha=6\Delta=2\pi/28$.  As a
base-level comparison we also compute an under-resolved (or no-model)
solution of the MHD equations at $N^3=168^3$.  Time evolution of the
total energies and enstrophy are shown in Fig. \ref{fig:sfs_global}.
In comparison with under-resolving MHD, LAMHD$-\alpha$ shows errors of
approximately the same magnitude in these global quantities.
Comparisons of energy spectra (Fig. \ref{fig:sfs_spectra}), however,
show an improvement in predictive quality for LAMHD$-\alpha$,
especially for the magnetic energy spectrum.  As turbulence develops,
energy begins to pile up at small-scales and deplete at intermediate
scales for $168^3$ MHD.  {LAMHD$-\alpha$ improves the prediction
  of superfilter-scale spectra compared to no SFS model.}

\begin{figure}[htbp]
\centering
  \includegraphics[width=0.95\linewidth]{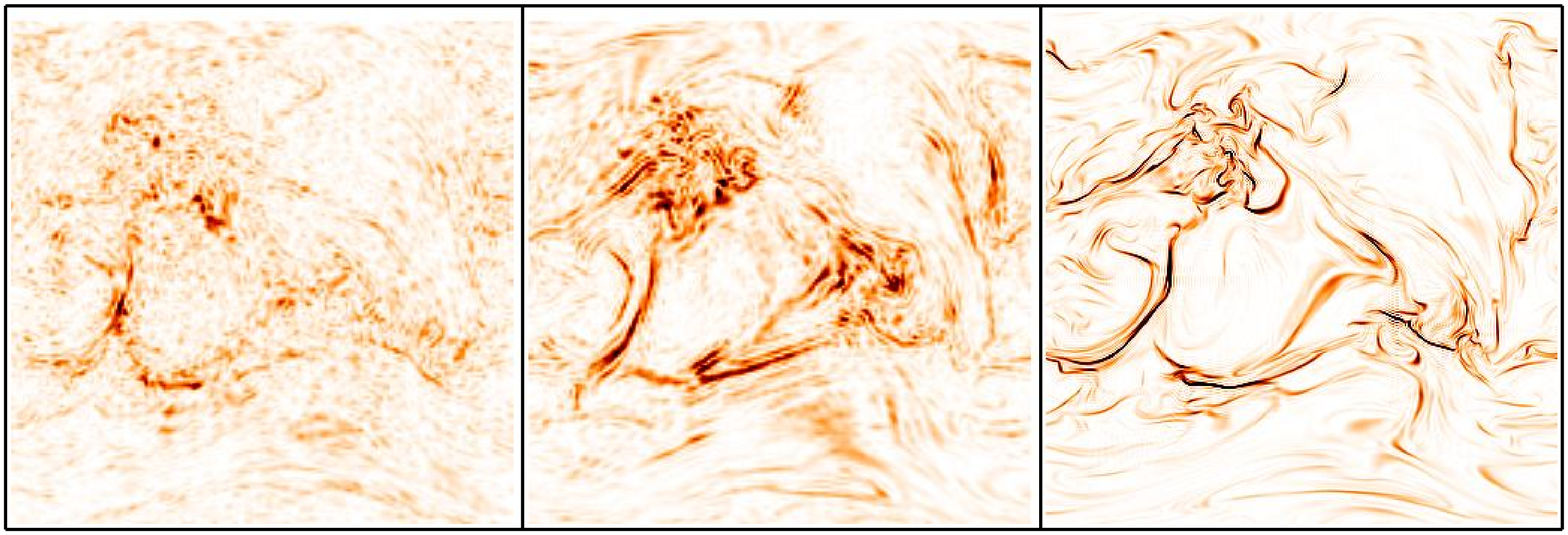}
  \caption{Cross sections of square current, $j^2$, at $t=8.4$ for
    no-model $168^3$ (left), $168^3$ LAMHD$-\alpha$ (center), and
    $1024^3$ DNS (right) at full resolution.  LAMHD$-\alpha$ provides a much
    better capturing of the intermittent current sheets than the
    under-resolved solution.}
  \label{fig:sfs_sheet}
\end{figure}

{Cross-sections of $|\vec{j}|^2$, shown in
  Fig. \ref{fig:sfs_sheet} at $t=8.4$ indicate that LAMHD-$\alpha$
  finds sharper and better defined, more intermittent current sheets
  than the under-resolved run compared to the DNS.}

\section{Summary}

Incompressible LANS$-\alpha$, while it performed well at moderate
Reynolds number is limited as a high $Re$ SFS model.  Due to its
strong suppression of spectrally local interactions at
subfilter-scales, and consistent with its conservation of small-scale
circulation, LANS$-\alpha$ develops rigid bodies which contaminate the
superfilter-scale energy spectrum.  In contrast, Clark$-\alpha$ and
Leray$-\alpha$, neither of which conserve small-scale circulation do
not develop energy-spectrum contamination from rigid bodies.
LANS$-\alpha$, however, best matches the intermittency properties of
Navier-Stokes {fluid turbulence.}

In MHD, {a} mechanism for local small-scale transfer is the interaction
of small-scale Alfv\'en waves.  As
LAMHD$-\alpha$ supports Alfv\'en waves at all scales
while slowing and hyperdiffusively damping those with wavelength $\lambda<\alpha$, it
more gently suppresses SFS local interactions than
LANS$-\alpha$. 
This together with the greater nonlocality in MHD and the Lorentz-force
source of small-scale circulation, {inhibits} the formation of rigid
bodies in LAMHD$-\alpha$.  \neu{It appears to retain the good intermittency properties of LANS$-\alpha$ without its poor spectral properties.} For this reason, we find LAMHD$-\alpha$ to
be a viable model at high $Re$ in 3D.  As LAMHD$-\alpha$ has been
previously found to reproduce the difficult to model properties of MHD
at high $Re$ in 2D \cite{MMP05a} and moderate $Re$ in 3D \cite{MMP05b}, we believe it will prove to be a generally applicable MHD LES, in many instances in geophysics and astrophysics where magnetic fields are known to be important dynamically.

%
%
%



\printindex
\end{document}